\documentclass[12pt,preprint]{aastex}

\newcommand{\Lya}{Ly$\alpha$}
\newcommand{\hal}{H$\alpha$}
\newcommand{\hbeta}{H$\beta$}

\newcommand{\cm}{cm$^{-2}$}
\newcommand{\kms}{km~s$^{-1}$}

\slugcomment{submitted to The Astrophysical Journal, Letters}

\shorttitle{\Lya\ Images of Local Starbursts}
\shortauthors{Kunth et al.}

\begin{document}


\title{The First Deep ACS \Lya\ Images of Local Starburst Galaxies}

\author{Daniel Kunth}
\affil{Institut d'Astrophysique, Paris, 98bis Bld Arago, F-75014, Paris,
France}
\email{kunth@iap.fr}

\author{Claus Leitherer}
\affil{Space Telescope Science Institute, 3700 San Martin Dr., Baltimore, MD
21218}
\email{leitherer@stsci.edu}

\author{J. Miguel Mas-Hesse\altaffilmark{1}}
\affil{Centro de Astrobiolog\'\i a (CSIC-INTA), E-28850 Torrej\'on de Ardoz,
Madrid, Spain}
\email{mm@laeff.esa.es}

\author{G\"oran \"Ostlin}
\affil{Stockholm Observatory, SE-106 91 Stockholm, Sweden}
\email{ostlin@astro.su.se}

\and

\author{Artashes Petrosian}
\affil{Byurakan Astrophysical Observatory and Isaac Newton
Institute of Chile, Armenian Branch, Byurakan 378433,
Armenia}
\email{artptrs@yahoo.com}

\altaffiltext{1}{Laboratorio de Astrof\'{\i}sica y F\'{\i}sica Fundamental,
POB 50727, E-28080 Madrid, Spain}

\vfill
\eject

\begin{abstract}
We report the first results from a deep \Lya\ imaging program of local
starburst galaxies with the Advanced Camera for Surveys (ACS) of the
Hubble Space Telescope. The two observed galaxies ESO~350-IG038 and
SBS~0335-052 have luminosities similar to those of the Magellanic Clouds
but differ in their chemical composition. ESO~350-IG038 has an
oxygen abundance of 1/8 solar, whereas SBS~0335-052 is known to have
one of the lowest abundances among blue galaxies ($\sim$1/30).  The ACS
imaging reveals a complex \Lya\ morphology, with sometimes strong
offsets between the emission of \Lya\ and the location of stellar
light, ionized gas traced by H$\alpha$, and the neutral gas. Overall,
more \Lya\ photons escape from the more metal- and dust-rich
galaxy ESO~350-IG038. The absence of clear SBS~0335-052 \Lya\ emission over
all
observed knots, whatever their dust content or/and color indices,
contradicts
  model expectations of a lower
escape fraction from dust-rich gas due to destruction of \Lya\
photons by dust grains. Rather, the results are in qualitative
agreement with models suggesting the kinematic properties of the gas
as the dominant \Lya\ escape regulator. If the properties of
the two observed galaxies are representative for starburst
galaxies in general, \Lya\ will be difficult to interpret as
a star-formation indicator, in particular if based on \Lya\ imaging at
low spatial resolution.
\end{abstract}




\section{Introduction}



\Lya\ is now widely used as a tracer of galaxy formation and evolution
in the high redshift universe.
 For several decades
many observers
have attempted to understand the conditions required to detect this line by
focusing on nearby regions of rapid star formation. In a recent
study,  Kunth et
al. (1998)  discussed a set of eight nearby star-forming galaxies. They
found four  \Lya\ emitting galaxies with clear \Lya\ P-Cygni profiles
indicating large-scale outflows of the interstellar medium (ISM).
In contrast, the four remaining
galaxies displayed damped \Lya\ absorption instead of emission, regardless
of
their dust content. Thuan \& Izotov (1997) studied the unique starburst
Tol1214-277 whose very strong and symmetric \Lya\ line shows no trace of
blueshifted absorption.

The Space Telescope Imaging Spectrograph (STIS) on board HST was used to
reobserve three of the galaxies discussed by Kunth et al. (1998): two with
a \Lya\ P-Cygni profile and a third with only a broad \Lya\ absorption.
 2-D analysis of these spectra clearly shows some partial decoupling of the
UV
continuum source and \Lya\ emission (Mas-Hesse et al. 2003). The spectra
confirm the previous GHRS observations that favor a large \Lya\
escape probability when most of the neutral gas is velocity-shifted
relative to the ionized regions and show that  expanding shells of
neutral gas can reach a few kpc wide.

These observational results are  well reproduced by recent
models of Tenorio-Tagle et al. (1999) for the evolution of a
superbubble surrounding a starburst in a host galaxy.  Whenever the
neutral gas is outflowing, the \Lya\ photons redward of 1216~\AA\ can
escape.  In contrast, no emission is detected if the ionized gas is
shielded by a homogeneous slab of static neutral gas with column density
above $N \approx 10^{18}$~\cm.  On the other hand, diffuse \Lya\ may still
be present and leaking out of the HI cloud, even if the HI is
homogeneous and static. Finally, if the starburst is very young or if the HI
coverage is static but porous, \Lya\ emission may be seen directly from the
HII region, but with an intensity weaker than the recombination value.
These results highlight the additional importance of the ISM properties for
the escape probability of \Lya: The galaxies observed to date span a
metallicity
range of more than a factor of 10 but display no correlation between metal
abundance and \Lya\ emission strength.

However additional parameters complicate the comparison between
theory and observations:
 \Lya\ photons escape may critically depend on the column density
and distribution of the neutral gas and dust, the morphology of the
supershells, and probably on the luminosity of the burst and on the
morphology of the host galaxy. In order to tackle these issues, a
pilot \Lya\ {\em imaging} survey of nearby star-forming galaxies has been
conducted using the Advanced Camera for Surveys (ACS) on board HST. An
initial sample of six local starburst galaxies was chosen for Cycle~11,
and here we report on initial results for the first two targets.

Throughout this paper we have assumed a value for the Hubble constant of
$H_0 = 75$~\kms

\section{ACS observations and data reduction}

30 orbits had been allocated to this program (GO-9470). Here we discuss the
results for the first two galaxies ESO\,350-IG038 and
SBS\,0335-052. Previous GHRS spectroscopy showed the first galaxy to be a
\Lya\ emitter while the second exhibited a broad damped absorption.
The observations were obtained with the Solar Blind Channel (SBC) of the
ACS.  Each galaxy was observed during five orbits in two adjacent filters.
The F122M filter contains the \Lya\ emission, while the adjacent F140LP
longpass does not transmit \Lya\ and was chosen for the continuum
subtraction.  The SBC uses a MAMA detector which is not subject to cosmic
rays, and one single exposure per orbit was obtained in each filter.  The
F122M images were obtained during the ``SHADOW'' part of the orbit in order
to lower the background from geocoronal \Lya\, while the F140LP ones were
acquired during the rest of the orbit. The total integration  times per
galaxy
in the F122M and F140LP filters were 9095 and 2700 s, respectively.
As a result of the larger bandwidth,
the F140LP observations have approximately twice the S/N of the F122M data.
The images were dithered using 3 positions with offsets of $\sim 10$ pixels
in between. The SBC utilizes a pixel scale of $0.032$\arcsec, and the
effective field of view is $31\arcsec \times 35$\arcsec\ (Pavlovsky et
al. 2002).

The ``drizzled'' images produced by the standard pipeline for the two
filters were combined after coalignment.  From each combined image, the
background was approximated by a surface (parameterized with a first-degree
polynomial), and subtracted.  Although weak in absolute terms in both
filters, the background
was significantly stronger in the F122M filter due to geocoronal emission.
The images, calibrated in counts per second, were multiplied by the
PHOTFLAM keyword  from the image headers and
converted to erg~s$^{-1}$~cm$^{-2}$~\AA$^{-1}$ at
the pivot wavelength of each filter (1274 and 1527~\AA, respectively).

Our ACS images were taken with the goal of detecting \Lya\ emission from
massive star-forming regions, and possible diffuse emission.  In order to
detect the excess emission within the F122M filter that would be produced
by the \Lya\ photons, we need to completely remove the continuum
contribution. The subtraction procedure depends on the slope of the
continuum, which might be variable for different regions over the
images. Therefore, different normalizations between both filters were
required for different regions.

We express the continuum spectral energy distribution as a power law:
$f_\lambda \propto \lambda^\beta$. For a flat continuum ($\beta=0$),
the scaling factor between F122M and F140LP is 10.3, i.e., the
ratio of the PHOTFLAM keywords. A redder SED with
$\beta=1$ gives a scaling factor of 12.5, and a bluer SED with $\beta=-2$
gives 7.2. We have used available spectroscopic UV
data from GHRS (Thuan \& Izotov 1997, Kunth et al. 1998) and IUE, as well as
the optical to UV colors from HST photometry to estimate $\beta$ and
aid us in the continuum subtraction.

%

We stress that the continuum subtraction procedure is affected by several
parameters, most importantly the intrinsic continuum slope of the stellar
population.
Moreover, a \Lya\ P-Cygni profile as frequently observed in
these objects tends to hide the \Lya\ emission when imaged through this
filter (the emission and the absorption wings tend to cancel, at least
partially). Therefore, our results should be regarded qualitatively only.

Our subtraction technique is supported by the comparison with previous GHRS
data. As discussed below, we find excess emission within
the F122M filter where GHRS shows a prominent \Lya\ emission line as well
(ESO350-IG038), while we detect a relative deficit of photons around \Lya\
where a strong and broad absorption profile is present (SBS0335-052).

%
%

\section{Results and discussion}

\subsection{ESO350-IG038 (Haro 11)}

This galaxy has a radial velocity of 6175~\kms\ and a complex
morphology.  The optical and near-infrared (IR) properties have been
discussed by Vader et al.  (1993), Heisler \& Vader (1994), Bergvall et
al. (2000), and Bergvall \& \"Ostlin (2002). The nebular oxygen abundance
is 12+log(O/H)=7.9 (Bergvall \& \"Ostlin 2002).
Three bright condensations
have been identified, all with strong \hal\ emission. Following the
nomenclature by Vader et.  al. (1993), we denote these by A, B, and C (see
Fig.~1).  Knot A has the bluest optical and near-IR colors and small
reddening, $E(B-V)= 0.16$, as derived from \hal/\hbeta . In contrast, B
and C have redder colors and larger internal extinction, $E(B-V)\approx
0.4$.
The galaxy has been imaged in the WFPC2/F606W filter in an HST
snapshot program (Malkan et al. 1998) which has revealed numerous compact
star clusters (\"Ostlin 2000).
A \Lya \ line exhibiting a P-Cygni profile was detected with
 HST/GHRS (Kunth et al. 1998).

Fig.~1 shows the F140LP, F122M, and  continuum subtracted \Lya \ images.
A relative scaling of 10.3 between F122M and F140LP
was used, equivalent to $\beta = 0$.
In Fig.~2 we show an RGB composite of optical,  F140LP, and continuum
subtracted \Lya \ images.


Figures 1 and 2 show that knot A is resolved into several bright star
clusters,
whereas B is more diffuse and C is dominated by a single luminous source.
Comparison of the F140LP and F606W images shows the regions around A and C
to
have very blue UV/optical colors:
$f_{\lambda,F140LP}/f_{\lambda,F606W} \approx 15$, implying $\beta=-2$,
whereas
in region B the average ratio
is only 3, implying $\beta=-0.8$. The region east of C is redder,
$f_{\lambda,F140LP}/f_{\lambda,F606W}
\approx 1$, as seen also from Fig. 2.


\Lya \ emission is detected from region C. It is more extended along the
north-south direction than the continuum, indicating that the emission
is not isotropic. \Lya \ is detected in region~A as well.
Faint \Lya \ emission is seen over a large fraction of
the galaxy, with the exception of region B and its immediate surroundings
showing clear absorption. The emission in regions A and C has
absorption cores near the knots. However, the absorption cores are shifted
slightly to the north with respect to the emission. We have confirmed that
this is not due to a misalignment of the F122M and F140LP images. After
integration,
the emission outweighs the absorption.
A faint \Lya\ blob (labeled `D' in Fig. 1), undetected in the continuum,
is seen  south-east of C.

The results are sensitive to the scaling of the two filters F122M and F140LP
when subtracting the continuum.
If we use the $f_{\lambda,F140LP}/f_{\lambda,F606W}$ ratio and assume
a power law spectral energy distribution (SED) we obtain $\beta=-2$ to 0.
Using this method to compute a spatially varying scaling
factor between F122M and F140LP, the effect is that only C and D shows
significant emission.
However, the GHRS spectrum of  Kunth et al. (1998) shows a continuum
decrease from 1300 to 1200~\AA, indicating that one power law SED is
not a good approximation over the full spectral range. Adopting the GHRS
slope still leads to absorption from region B. Using the slope derived from
an IUE spectrum ($\beta=-1$) gives a result similar to that presented in Fig
1;
the emission close to A becomes weaker, but the diffuse emission remains.
This illustrates the importance of determining the right continuum level.
However, given the appearance of the continuum near \Lya \ from the GHRS
spectra, it is clear that photometry always faces a potential risk of over-
or underestimate the \Lya \ line strength.

The absorption in region B, and in the cores of the sources in A and C is
too strong to be explainable by anything but a deep \Lya \ through. Note
that B still has rather blue UV to optical colors, so the absorption cannot
be a spurious effect of reddening or an underlying red stellar population.

Double velocity components in \hal\ with a velocity difference of
$\sim 50$~\kms\ are seen in regions A, C, as well as B (\"Ostlin et
al. 1999).  Kunth et al. (1998) detected broad metallic absorption lines
from neutral gas blue-shifted by 60~\kms\ with respect to \Lya .
At variance with
SBS0335-052, no HI emission gas has been detected at a limit of $10^8
M_\odot$ (Bergvall et al. 2000), although Kunth et al. (1998) detected
dense neutral gas in absorption.
Hence the neutral gas distribution is likely to be clumpy, and any
diffuse gas would be ionized, explaining why we detect \Lya \ in emission.

The small but systematic offset of \Lya\ to the south with respect to the
continuum in regions A and C might be an indication of a velocity gradient
in the neutral gas. This is consistent with the \hal\ velocity field
showing a global velocity gradient of the order of $10-15$~km~s$^{-1}$ per
arcsec,
or $25-40$~km~s$^{-1}$~kpc$^{-1}$ (uncorrected for inclination) with a
kinematical major
axis in the south-eastern direction, causing a
blueshift of the gas to the south of A and C with respect to these regions
(\"Ostlin et al. 1999).
The redshifted regions to the north of A and C do not show \Lya\ but
these photons may be absorbed by the same gas complex responsible for the
absorption in the center (region B).
As discussed in detail in Mas-Hesse et al. (2003), the visibility of the
\Lya\
emission increases with the blueshift of the neutral
gas. While we expect \Lya\ photons to be escaping from the regions where
the neutral gas is blueshifted, we do not expect emission in the absence
of a velocity offset between the neutral and ionized regions.
The absence of emission from region B
might therefore indicate  static neutral gas along our sight line.  This
galaxy
illustrates the importance of the kinematic structure of the region for
the emission of \Lya\ photons. The resulting \Lya\ distribution will be
non-isotropic and much more complex than the \hal\ distribution.

As mentioned above, the \Lya\ emitting regions show absorption cores.
We attribute this effect to the known P-Cygni properties of the \Lya\
profiles in this galaxy: the stronger the
stellar continuum, the deeper will be the blueshifted absorption wing (see
Mas-Hesse et al. 2003 for details). Therefore, the emission wing
tends to cancel with the absorption where the continuum is higher.

\subsection{SBS0335-052}
SBS0335-052 has a nebular oxygen abundance (12 + log(O/H) = 7.3; Melnick et
al.
1992). This value is almost as low as that of IZw18, triggering discussions
whether SBS0335-052 is
a truly young galaxy in its first burst of star formation
(Thuan et al. 1997, cf. \"Ostlin \& Kunth 2001).
The broad damped \Lya\ absorption reported by Thuan et al. (1997) from
their GHRS spectra suggests a static HI reservoir surrounding the massive
star condensations. The HI column density within the GHRS aperture amounts
to $7.0\times 10^{21}$ cm$^{-2}$. According to Mas-Hesse et al. (2003), the
very young starburst episode would be starting to ionize the surrounding
medium, but the mechanical energy released would still be insufficient
to accelerate the large amount of HI gas in front of the region. The fits
presented by these authors indicate that the intrinsic \Lya\ emission line
could be very strong, with an equivalent width around 120 \AA, as predicted
by theoretical models. HST WFPC2 V and I images (Thuan, Izotov, \&
Lipovetsky
1997), UBVRI surface photometry by Papaderos et al. (1998), and near- and
mid-IR imaging (Vanzi et al. 2000; Dale et al. 2001) have revealed several
super star clusters (SSC).  A comparison of the data of Thuan et al. with
our ACS frames (Figs. 1 and 2) shows that the brightest star-forming knots
in the visible and IR range (SSC1 and SSC2 in the notation of Thuan et al.)
are not the brightest in the UV continuum.  In contrast, SSC4 and SSC5 are
much brighter in the UV than the other SSCs:
$f_{\lambda, F140LP}/f_{\lambda, F791W} \approx 100, 100$, and 35 for
SSC4, SSC5, and SSC1, respectively.  Our ACS images show that SSC3
is a well defined double system which is not resolved in the optical and
near-IR.

A clear lack of photons  within the \Lya\ filter is evident for the
different SSCs on the image.
This lack of photons is intrinsic, and is not related to any normalization
effect. This is remarkable in view of the strong \hal\ observed in this
galaxy
(e.g. Melnick et al. 1992) and the $f_{\lambda, F140LP}/f_{\lambda, F791W}$
colors.
This absence is found both when the intrinsic slope of the UV continuum is
used ($\beta=-2.5$) and also when a flat continuum ($\beta=0$) is assumed.
If  $\beta$ is estimated from $f_{\lambda, F140LP}/f_{\lambda, F791W}$, it
is found to vary between --3 and --1.8 over the starburst region.
Nevertheless, a weak and extended diffuse \Lya\ emission
is detected around SSC5 to the N-E, as shown in Fig.~1 (bottom
right). This emission is very weak, and its reality is subject to
 the uncertainties in the
continuum subtraction.
\Lya\ photons could escape if they find a way: either through holes
with lower HI column density, or regions where the neutral gas could be
moving at some velocity with respect to the central HII region.
 Tenorio-Tagle et al. (1999) (see also
Mas-Hesse et al. 2003) suggest that  an ionized cone could develop
through the surrounding
neutral gas halo shortly after the ignition of a
massive starburst. This cone would be oriented along the minor axis of the
system. In this case, most of the gas towards the N-E of SSC5 could be
already ionized, allowing some \Lya\ photons to escape.  Thuan
\& Izotov (1997) found evidence for neutral gas flows,
identifying two systems at $-500$ and $-1500$ km s$^{-1}$ with respect to
the galaxy redshift. As concluded for ESO 350-IG038, the kinematic structure
of
the neutral gas in these regions could favor the escape of some \Lya\
photons in specific areas. Long-slit high-resolution
spectroscopy would be needed to identify the dominant mechanism leading
to the detection of \Lya\ photons.

In any case, the absence of clear net \Lya\ emission from all observed
knots confirms
what has already been discussed (Thuan et al. 1997; Kunth et al. 1998),
supporting again the lack of correlation between \Lya\ fluxes and
metallicities (hence dust) found by Giavalisco et al. (1996).
 The large aperture of the IUE satellite limited Giavalisco et al. 
to size scales of order kpc. Our highly resolved ACS images allow us 
to push the  1-pixel resolution down to 5.7 and 8.9 pc for ESO350-38 
and SBS0335-052, respectively. If dust affects the \Lya\ escape, 
it must do so at even smaller spatial scales. While some
dust is clearly present in SBS0335-052, SSC1 does not suffer from severe
extinction (Thuan et al. 1997), and the same must be true for SSCs 4 and 5
in view of their UV to optical colors. Even if the emission close to
SSC5 is real, it is very weak and completely negligible on the whole.
 The HI cloud around SBS0335-052 is clearly
very dense and apparently static (at least in front of the SSCs), hence
favoring the multiple scattering mechanism responsible for the
attenuation.

\section{Conclusions and implications for high-redshift galaxies}

Our first ACS images of two dwarf starburst galaxies give us a view at the
\Lya\ emission that is complementary to that of spectroscopy. These images
already confirm the existence of two markedly different types of starburst
galaxies: some are \Lya\ emitters, while others show rather broad
absorption profiles. Even galaxies displaying \Lya\ emission may do so in
a complex pattern, as we have shown in ESO350-38: two bright \hal\ regions
show emission, while a third one shows strong absorption. At high
redshift, this morphological resolution is not attainable.

In many respects, \Lya\ could be a fundamental probe of the young universe.
It suffers from fewer luminosity biases than Lyman-break techniques so that
\Lya\ surveys become a more efficient way to trace the fainter end of the
luminosity function, i.e., it traces the building blocks of present-day
galaxies in the hierarchical galaxy formation paradigm (Hu, Cowie, \&
McMahon 1998; Fynbo et al. 2001). Early results paint a complex
picture. The equivalent widths of the sources are much larger than expected
for ordinary stellar populations (Malhotra \& Rhoads 2002). They could be
explained by postulating an initial mass function (IMF) biased towards more
massive stars, as predicted theoretically for a very metal-poor stellar
population (Bromm, Coppi, \& Larson 2001).  The combined effect of low
metallicity and flat IMF, however, can only partly explain the anomalous
equivalent widths. Additional mechanisms must be at work. Could  spatial
offsets between the escaping
\Lya\ and the stellar light, together with higher extinction of the dust
shrouded stars be important?
Our ACS imagery cautions against using the \Lya\ equivalent width as
a star-formation indicator in the absence of spatial information.



The source numbers themselves are only about 10\% of the numbers expected
from an extrapolation of the Lyman-break luminosity function. Malhotra \&
Rhoads (2002) speculate if the youngest galaxies are preferentially
selected, whereas older populations are excluded, the results are skewed
towards large \Lya\ equivalent widths.  Could dust formation after
$10^7$~yr destroy the \Lya\ photons? The results in the low-redshift
universe suggests otherwise. We find no support for dust playing a {\em
major} role in destroying \Lya\ photons. Rather, the ACS images favor
 the highly irregular gas morphology as key to
 understanding the \Lya\ escape mechanism via kinematic effects.

Haiman \& Spaans (1999) proposed \Lya\ galaxies as a direct and robust test
of the reionization epoch. Prior to reionization, these galaxies are hidden
by scattering of the neutral intergalactic medium (IGM). Therefore, a
pronounced decrease in the number counts of {\em galaxies} should occur at
the reionization redshift, independent of Gunn-Peterson trough observations
using {\em quasars}. \Lya\ in galaxies would have the additional appeal of
being sensitive at much higher IGM optical depths since its red {\em wing}
coincides with the red damping wing of intergalactic \Lya\ only. While this
idea is attractive in principle, our ACS imagery calls for caution. In
practice, \Lya\ is a complex superposition of emission and absorption in
the star-forming galaxy itself.  The resulting \Lya\ profile will be
strongly affected by absorption from the continuum. This becomes even more
of a concern when imaging data are interpreted.  When integrating over the
filter bandpass, the emission part of the profile is partly compensated by
the absorption.  As a result, we measure significantly lower \Lya\ fluxes
than with spectroscopic methods, and the escape fraction of \Lya\ photons
is significantly underestimated. Finally it must be kept in mind that
in absence of any prior spectroscopic information on the continuum
distribution, the required continuum subtraction could be completely
erroneous.

The \Lya\ line is a premier star-formation tracer, in particular at high $z$
where traditional methods, such as H$\alpha$ or the far-IR emission,
become impractical. Radiative transfer effects in the surrounding
interstellar gas make its interpretation in terms of star-formation
rates less straightforward than often assumed. Clearly,
a better understanding of the complex \Lya\ escape mechanisms, both
empirically and theoretically, is required before we can attempt
to interpret large-scale \Lya\ surveys. 

\acknowledgments

A.R.P.acknowledges the hospitality of the Space Telescope Science Institute
during his stay as visiting scientist supported by the Director's
Discretionary Research Fund.
This work was supported by HST grant GO-9470.01-A from the Space
Telescope Science Institute, which is operated by the
Association of Universities for Research in Astronomy, Inc.,
under NASA contract NAS5-26555. JMMH has been partially supported by
Spanish grant AYA2001-3939-C03-02. G\"O acknowledges support from
the Swedish research council and the Swedish national space board.

\clearpage

\begin{figure}
\epsscale{0.6}
\plotone{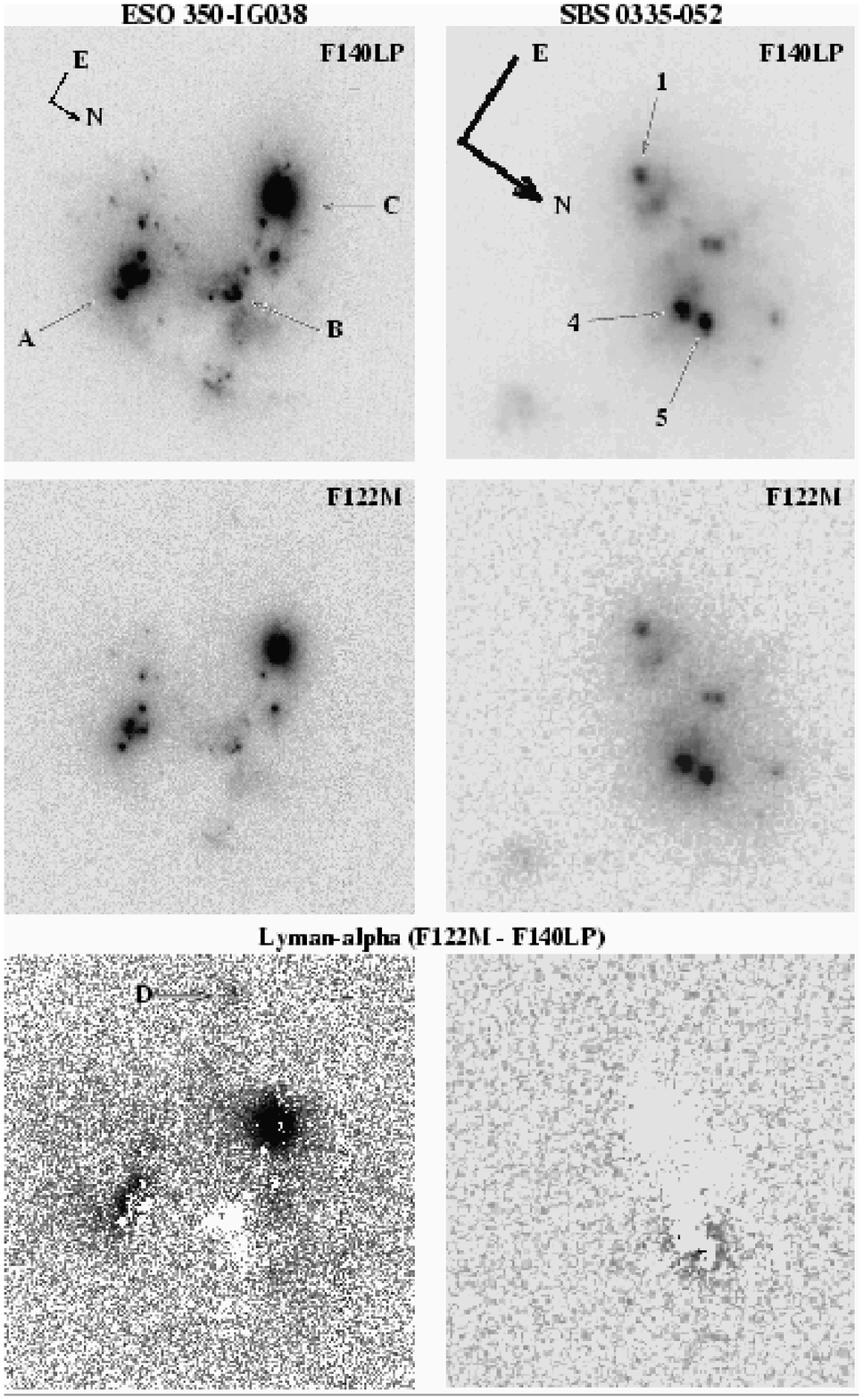}

\caption{HST/ACS images (logarithmic intensity scaling).
Top left: UV continuum (F140LP) image for ESO\,350-IG038 with the
orientation indicated in the upper left corner. The field shown is
$13\arcsec \times 13\arcsec$. The three main emission regions discussed in
the text are labeled A, B and C.  Middle left: F122M image of
ESO\,350-IG038.  Lower left: continuum subtracted \Lya\ image using a
relative scaling of 10.3 between F140LP and F122M. Note the faint \Lya\
emitting blob (labeled D) which is not seen in the continuum image.  Top
right: UV continuum (F140LP) image for SBS\,0335-052. The field shown is
$4.4\arcsec \times 4.4\arcsec$. The star forming regions discussed in the
text are labeled '1','4' and '5'.  Middle right: F122M image of
SBS\,0335-052.  Lower right : continuum subtracted \Lya\ image. Note the
faint emission to the north of '4' and '5'.  }
\label{fig1}
\end{figure}

\clearpage

\begin{figure}
\epsscale{0.6}
\plotone{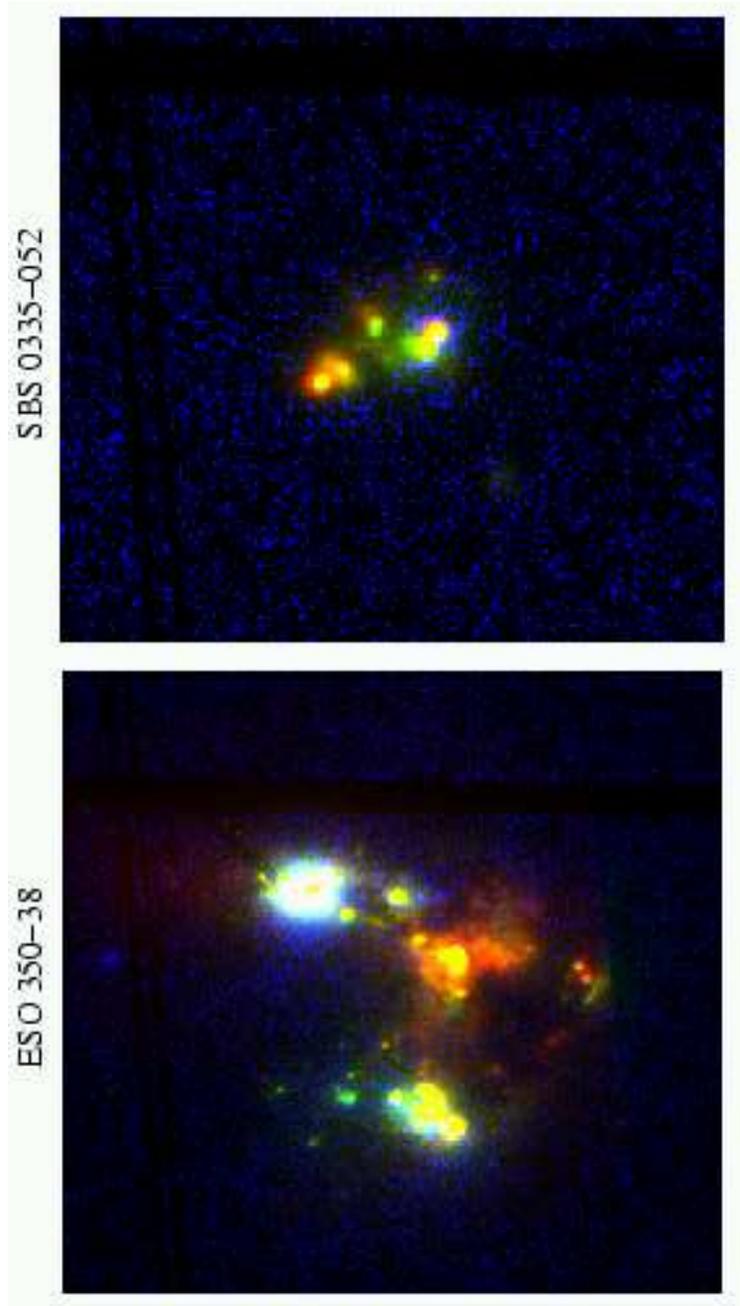}
\caption{Color composite images of the two galaxies, with the optical image
coded in red, the F140LP in green and the continuum subtracted \Lya\ in
blue.  The spatial scales have been chosen such that both images correspond
to the same physical size ($2.3 \ {\rm kpc} \times 2.3 \ {\rm kpc}$ for
$H_0 = 75$~\kms~Mpc$^{-1}$).  Left: ESO\,350-38, with WFPC2/F606W in the
red channel. The size is $13\arcsec \times 13\arcsec$.
Right: SBS\,0335-052, with WFPC2/F791W (taken from the HST archive) in the
red channel. The size is $8.3\arcsec \times 8.3\arcsec$.}
\label{fig2}
\end{figure}

\clearpage

\end{document}